\documentclass[pre,aps,twocolumn,superscriptaddress,longbibliography]{revtex4-2}

\usepackage[english]{babel}
\usepackage{graphicx}
\usepackage{amssymb,amsfonts,amsmath}
\usepackage{color}
\usepackage[hidelinks]{hyperref}

\usepackage{tikz}
\usepackage{tikz-feynman}
\usetikzlibrary{calc}
\usetikzlibrary{positioning}
\usetikzlibrary{decorations.pathmorphing}
\tikzset{snake it/.style={decorate, decoration=snake}}

\newcommand{\MayerA}{
   \mayerdiagram{
   \node (a) at (0,0) [particle] {};
   \node (b) at (0.6,0) [particle] {};
   \draw[fbond] (a) -- (b);
}
}

\newcommand{\MayerB}{
   \mayerdiagram{
     \node (A) at (0,-0.2) [particle] {};
     \node (B) at (0.5,-0.2) [particle] {};
     \node (C) at (0.25,0.2) [particlefilled] {};
     \draw[fbond] (A)--(B);
     \draw[fbond] (B)--(C);
     \draw[fbond] (C)--(A);
   }
}

\newcommand{\MayerC}{
   \mayerdiagram{
     \node (A) at (0,-0.25) [particle] {};
     \node (B) at (0.5,-0.25) [particle] {};
     \node (C) at (0,0.25) [particlefilled] {};
     \node (D) at (0.5,0.25) [particlefilled] {};
     \draw[fbond] (A)--(B);
     \draw[fbond] (B)--(D);
     \draw[fbond] (C)--(D);
     \draw[fbond] (C)--(A);
   }
}

\newcommand{\MayerD}{
   \mayerdiagram{
     \node (A) at (0,-0.25) [particle] {};
     \node (B) at (0.5,-0.25) [particle] {};
     \node (C) at (0,0.25) [particlefilled] {};
     \node (D) at (0.5,0.25) [particlefilled] {};
     \draw[fbond] (A)--(C);
     \draw[fbond] (A)--(D);
     \draw[fbond] (B)--(C);
     \draw[fbond] (B)--(D);
   }
}

\newcommand{\MayerE}{
   \mayerdiagram{
     \node (A) at (0,-0.25) [particle] {};
     \node (B) at (0.5,-0.25) [particle] {};
     \node (C) at (0,0.25) [particlefilled] {};
     \node (D) at (0.5,0.25) [particlefilled] {};
     \draw[fbond] (A)--(C);
     \draw[fbond] (A)--(D);
     \draw[fbond] (B)--(C);
     \draw[fbond] (B)--(D);
     \draw[fbond] (A)--(B);
   }
}

\newcommand{\MayerF}{
   \mayerdiagram{
     \node (A) at (0,-0.25) [particle] {};
     \node (B) at (0.5,-0.25) [particle] {};
     \node (C) at (0,0.25) [particlefilled] {};
     \node (D) at (0.5,0.25) [particlefilled] {};
     \draw[fbond] (A)--(C);
     \draw[fbond] (A)--(D);
     \draw[fbond] (B)--(C);
     \draw[fbond] (B)--(D);
     \draw[fbond] (C)--(D);
   }
}

\newcommand{\MayerG}{
   \mayerdiagram{
     \node (A) at (0,-0.25) [particle] {};
     \node (B) at (0.5,-0.25) [particle] {};
     \node (C) at (0,0.25) [particlefilled] {};
     \node (D) at (0.5,0.25) [particlefilled] {};
     \draw[fbond] (A)--(C);
     \draw[fbond] (A)--(D);
     \draw[fbond] (B)--(C);
     \draw[fbond] (B)--(D);
     \draw[fbond] (C)--(D);
     \draw[fbond] (A)--(B);
   }
}

\newcommand{\MayerH}{
   \mayerdiagram{
     \node (A) at (0,-0.25) [particle] {};
     \node (B) at (0.5,-0.25) [particle] {};
     \node (C) at (0,0.25) [particlefilled] {};
     \node (D) at (0.5,0.25) [particlefilled] {};
     \draw[fbond] (A)--(C);
     \draw[fbond] (A)--(D);
     \draw[fbond] (B)--(D);
     \draw[fbond] (C)--(D);
     \draw[fbond] (A)--(B);
   }
}

\newcommand{\MayerSpecial}{
   \mayerdiagram{
     \node (A) at (0,-0.25) [particle] {};
     \node (B) at (0.5,-0.25) [particle] {};
     \node (C) at (0,0.25) [particlefilled] {};
     \node (D) at (0.5,0.25) [particlefilled] {};
     \draw[fbond] (A)--(C);
     \draw[fbond] (A)--(D);
     \draw[fbond] (B)--(C);
     \draw[fbond] (B)--(D);
     \draw[ultra thick,densely dotted] (C)--(D);
   }
}

\usepackage{tikz}
\usetikzlibrary{shapes}
\usetikzlibrary{patterns}
\usetikzlibrary{angles, quotes}
\definecolor{bostonuniversityred}{rgb}{0.8, 0.0, 0.0}
\definecolor{chromeyellow}{rgb}{1.0, 0.65, 0.0}
\tikzset{
   particle/.style={circle, draw, inner sep=1.5pt},
   particlefilled/.style={circle, draw, fill=black, inner sep=1.5pt},
   fbond/.style={thick},
   hbond/.style={thick, dashed},
}

\newcommand{\mayerdiagram}[1]{\tikz[baseline=-0.6ex]{#1}}
\newcommand{\rv}{{\mathbf r}}
\renewcommand{\vec}{\mathbf}

\newcommand{\avN}{\langle N\rangle}

\usepackage{amssymb}

\usepackage[final]{showlabels} 

\begin{document}


\title{
Hard disks confined within a narrow channel
}

\author{J.~M. Brader}
\affiliation{Department of Physics, University of Fribourg, CH-1700 Fribourg, Switzerland}
\author{E. Di Bernardo}
\affiliation{Department of Physics, University of Fribourg, CH-1700 Fribourg, Switzerland}
\author{S.~M. Tschopp}
\affiliation{Department of Physics, University of Fribourg, CH-1700 Fribourg, Switzerland}
\date{\today}

\begin{abstract}
We employ inhomogeneous integral equation theory to investigate the equilibrium properties of hard disks confined to a channel of width $L$ by hard parallel walls.
If the channel width is narrowed below 
two disk diameters, then the system enters a quasi one-dimensional regime for which the particles cannot move past each 
other.
In the limit when $L$ is equal to one particle diameter the system reduces to the one-dimensional bulk along the center of the channel.
We study first the dimensional crossover properties 
of the inhomogeneous Percus-Yevick (PY) integral equation as $L$ 
is reduced and then investigate the behaviour of a quasi one-dimensional system as the packing of the particles is increased 
for a fixed value of $L$. 
We find that the inhomogeneous PY equation is highly accurate for situations of quasi one-dimensional confinement and that it predicts 
the onset of a structural transition to a zigzag state at higher packing. 
The excellent performance of this integral equation method and 
the ease with which it handles confinement-induced dimensional 
crossover is a consequence of the improved resolution which comes from 
treating explicitly the inhomogeneous two-body correlation functions.
\end{abstract}

\maketitle

\section{Introduction}

One of the most effective theoretical techniques for calculating the thermodynamics and pair correlation functions of classical fluids in equilibrium is the method of integral equations based on approximate closures of the Ornstein–Zernike (OZ) equation \cite{hansen06,caccamo,JanssenIntegral}. For bulk fluid states this method provides quantitative first-principles predictions for a wide range of model interaction potentials. 
However, challenges remain in the vicinity 
of critical points, where fluctuations become important 
\cite{BraderThermophysics},  
and for high density state-points approaching the freezing transition. 
Indeed, a fundamental and long-standing question in the theory of fluids is whether the integral equation method is capable of indicating the existence 
of a bulk fluid-solid transition \cite{Kirkwood_freezing,Kozak_freezing}.

Extension of the integral equation approach to treat 
inhomogeneous systems in external potential fields is, in principle, quite straightforward. 
Tried and trusted closure approximations can be easily generalized to the inhomogeneous case, which enables the large body of empirical knowledge and experience gained from decades of research on bulk fluids to be exploited  
\cite{caccamo,JanssenIntegral}. 
However, inhomogeneous calculations require an additional exact sum-rule connecting the inhomogeneous two-body correlation functions to the spatially varying 
one-body density profile.
If the closure relation were exact, then the choice of sum-rule would become purely a matter of computational convenience, since identical results would be obtained in all cases. 
When employing an approximate closure, which we are typically compelled to do for models of interest, then the one-body density and two-body correlation functions predicted by the theory acquire a sum-rule dependence \cite{DouglasHenderson92,tschopp8,tschopp9}. 
Two popular choices for this sum-rule are the expressions derived by Lovett, Mou, Buff and Wertheim (LMBW) 
\cite{LMBW1,LMBW2,Lovett91,Lovett92,widom} 
and Yvon, Born and Green (YBG) 
\cite{Yvon_YBG,BornGreen_YBG,Lovett92,widom}.

The majority of studies employing the inhomogeneous integral equation method concern hard spheres in external fields with either planar or spherical symmetry. 
In these cases the closure of choice is typically the inhomogeneous generalization of the Percus-Yevick (PY) theory \cite{PercusOriginal}. 
This approximation is relatively easy to implement and is 
known to perform well for 
strongly repulsive interaction potentials. 
Predictions for the density distribution of fluids either around 
a solute particle or strongly confined within a slit are in excellent agreement with data from stochastic simulations 
\cite{AttardSpherical,AttardSpherical2,AttardBook,
KjellanderSarman1,KjellanderSarman2,nygard1} 
and even experiment \cite{nygard2}. 
Regarding crystallization, it was shown in \cite{Brader2008Struc-885} that the inhomogeneous integral equation approach can predict 
structural precursors to freezing within the dense hard-sphere fluid, thus demonstrating the ability of the method to detect the emergence of subtle microstructural ordering 
as the system becomes increasingly packed. 
The liquid-vapour interface of a phase separated Lennard-Jones system was addressed in \cite{Kovalenko} where the predicted binodal curve showed very good agreement with simulation data. 
Remarkably, the density profiles through the interface and the associated surface tension were found to be consistent with non-classical (i.e.~non-mean-field) critical exponents.   
Recent work has reinterpreted the inhomogeneous integral equation 
method in a way which makes it clear that this is a 
classical density functional theory (DFT) formulated on the level of the two-body correlation functions \cite{tschopp8,tschopp9}.

Since the pioneering work of Evans, DFT has established itself as the primary method for treating the statistical mechanics of classical many-body systems in external potentials \cite{evans79}. 
The central object of interest, at least within the standard implementation of the theory, is the grand potential functional. If this is given, then variational minimization generates complete information about all 
thermodynamic and microstructural properties of the inhomogeneous fluid. 
The practical drawback, however, is that the part of the grand potential functional dealing with 
interparticle interactions, the excess Helmholtz free energy functional, is usually not known exactly and accurate approximations exist for only very few model systems. 
An approach which has proved fruitful in constructing approximate 
functionals is to use dimensional crossover as a guiding principle \cite{RosenfeldTarazonaQuasi,TarazonaRosenfeld_0D,
Tarazona_Freezing,lutsko_0D}. 
The idea is that the application of external potentials which restrict the motion of the particles can be used to change the effective dimensionality of the system. Free energy 
functionals in either 2D or 3D (usually the cases of interest) 
should then recover any exact results known in either 1D or 
0D upon dimensional reduction. 
If approximate functionals which obey these limits could be obtained then they would be 
effectively hard-wired to avoid divergences or other pathological behaviours which could emerge under situations of strong confinement. 
However, in practice, obtaining explicit functional forms with the desired crossover properties is a difficult task. 
For fundamental geometrical reasons 
the case of 2D has proven to be particularly challenging; even the most sophisicated functional currently available is not capable of reducing to the correct 1D limit \cite{OettelDisk}.
Furthermore, investigations of dimensional crossover have so far focused exclusively on functionals for systems with hard-repulsive interactions. 
The crossover properties of functionals designed 
to treat systems with a soft repulsion or an attractive component to the pair interaction remain essentially unexplored. 

An interesting intermediate regime between 1D and 2D occurs when the particles are constrained such that they cannot pass each other (i.e.~strict longitudinal ordering is imposed), but still  
retain some amount of lateral motion. 
This quasi-1D situation can be realized, for example, by 3D spherical particles confined within a tube of small radius \cite{Q1D_Percus,Santos_tube} 
(relevant for microfluidics) or by disks in 2D constrained to move along a narrow channel \cite{Q1D_santos_eos,Q1D_santos_structure}. 
Research into these `almost 1D systems' was initiated in the early 1960s by Barker \cite{Q1D_Barker1,Q1D_Barker2} and subsequently led to the gradual development of a diverse toolbox of specialized techniques for exact analysis  \cite{Q1D_methods_virial,Q1D_methods_transfer,
Q1D_virial_singularity,Q1D_canonical_partition1,
Q1D_canonical_partition2}.
Indeed, for systems with hard repulsive interactions, these methods allow for exact computation of all thermodynamic and structural properties, a fact which has been largely (but not entirely \cite{RosenfeldTarazonaQuasi}), overlooked in discussions of dimensional crossover. 
Quasi-1D systems are of special value, since they provide a useful and instructive bridge 
between research into the exact solution of 1D models
and approximate DFT approaches to treating inhomogeneous 
fluids in higher dimensions.  
In addition to their intrinsic interest, the study of quasi-1D offers a simplified setting or test-bed, which can be used to gain insight 
into the microscopic mechanisms underlying both the freezing transition \cite{Q1D_kosterlitz,Q1D_kosterlitz_comment,Q1D_kosterlitz_reply} and the structure of amorphous packings in higher dimensional systems 
\cite{Q1D_glass1,Q1D_glass2}. 

In this paper we will employ the inhomogeneous integral equation method, consisting of the PY closure and the LMBW sum-rule \cite{tschopp9},
to investigate the properties of 
2D hard disks confined to a channel between two parallel hard walls. 
Our method provides direct access to both the one-body density profile and the inhomogeneous two-body correlation functions. 
Analysis of the latter provides detailed information about the 
internal microstructure, including correct prediction of the 
onset of long-range longitudinal order when the system is highly packed. 
The chosen model system has the advantage that it can be  
solved exactly within the quasi-1D regime \cite{Q1D_santos_eos,Q1D_santos_structure}, thus providing both a benchmark against which to test the predictions of the 
PY closure and a useful check on the quality of our numerical implementation.
We begin by investigating the dimensional crossover properties of the inhomogeneous PY theory as the channel width is reduced 
towards a single particle diameter. 
We show that, in contrast to standard DFT approaches based on an approximate free energy functional, our theory exhibits a natural crossover from 2D to the exact 1D solution \cite{percus_hard_rods} with no requirement for fine-tuning or consideration of specific 0D geometrical situations. 
We then proceed to study the evolution of the one-body density and inhomogeneous two-body correlation functions as a  
given quasi-1D channel becomes increasingly packed and observe the onset of long-range longitudinal order.

\section{Methods}

\subsection{The Ornstein-Zernike equation}
For bulk systems the two-body correlation functions are 
translationally and rotationally invariant and thus depend only 
upon the separation between the particle coordinates, $r_{12} \!\equiv\! |\vec{r}_1-\vec{r}_2|$. 
Given the total correlation function, $h$, and bulk density, $\rho_{\text{b}}$, we can define the two-body direct correlation 
function, $c$, using the OZ equation \cite{hansen06,mcquarrie}
\begin{equation}\label{bulk OZ equation}
h(r_{12}) = c(r_{12}) + \rho_{\text{b}}\! \int \! d\vec{r}_3 \, h(r_{13})\, c(r_{32}). 
\end{equation}
A closed system of equations can be obtained by introducing a 
supplementary (and usually approximate) relation between $h$ and 
$c$. 

In the presence of an external potential, $V_{\text{ext}}$, 
the symmetry of the bulk no longer holds. 
The one-body density then becomes a function of position, $\rho(\vec{r})$, and the two-body correlation functions become dependent on 
two vector arguments, e.g.~$h(r_{12})\!\rightarrow\!h(\rv_1,\rv_2)$. 
In this more general case the OZ equation takes the following form
\begin{equation} \label{Inhomogeneous OZ equation}
h(\rv_1,\rv_2) = c(\rv_1,\rv_2) 
+ \!\int\!\! d\rv_3\, h(\rv_1,\rv_3)\,  \rho(\rv_3) \, c(\rv_3,\rv_2),
\end{equation}
which reduces to equation \eqref{bulk OZ equation} in the absence of an external potential. 
Since equation \eqref{Inhomogeneous OZ equation} involves three unknown functions we require, in addition to a relation between 
$h$ and $c$, a further expression which connects the one-body density with the two-body correlations. 
For the purpose of the present work we will use the exact 
LMBW equation
\cite{LMBW1,LMBW2,Lovett91,Lovett92,widom}, given by
\begin{align}\label{LMBW}
\nabla_{\vec{r}_1}  \rho(\vec{r}_1) &=  -\rho(\vec{r}_1) \nabla_{\vec{r}_1} \beta V_{\text{ext}}(\vec{r}_1) \\
&+ 
\rho(\vec{r}_1)\!\int\! d \vec{r}_2 \,  c(\vec{r}_1,\vec{r}_2) \nabla_{\vec{r}_2} 
\rho(\vec{r}_2). \notag
\end{align}
The optimal choice of closure relation is often dependent upon the character of the pair interaction potential under investigation. 
A large body of semi-empirical knowledge about the best choice for any given situation has been accumulated from decades of research on the bulk OZ equation. 
Since we will later focus on two-dimensional systems of hard disks 
we choose to employ the inhomogeneous generalization of the 
PY approximation, which is known to be accurate for 
hard-particle interactions. 
For any pairwise interaction potential, $\phi$, this takes the following form
\begin{equation}\label{PY closure}
c(\vec{r}_1, \vec{r}_2) \!=\! 
\left( e^{-\beta \phi(r_{12})}\!-\!1 \right) 
\big(h(\vec{r}_1, \vec{r}_2) - c(\vec{r}_1, \vec{r}_2) + 1
\big)\,.
\end{equation} 
In the present work we will focus on the case of `hard spheres' 
in D-dimensions (i.e.~hard rods on a line in 1D, hard disks in 
a plane in 2D and hard spheres in 3D).
The expression \eqref{PY closure} is then equivalent to the 
more familiar PY conditions 
\begin{align}\label{HD PY closure}
h(\vec{r}_1, \vec{r}_2) &= -1, \qquad r_{12}<d ,\notag\\
c(\vec{r}_1, \vec{r}_2) &= 0, \qquad \;\;\; r_{12}>d,
\end{align}
where the particle diameter, $d$, will henceforth be set equal to unity.
Note that for hard rods in 1D the PY closure is exact.

We observe that the coupled pair of equations \eqref{Inhomogeneous OZ equation} and \eqref{PY closure} constitute a closed system for 
determination of the two-body correlation functions for any input one-body density. The two-body correlations are thus functionals 
of the one-body density, namely 
\begin{align}
\label{hfunctional}
h(\rv_1,\rv_2)&=h(\rv_1,\rv_2;[\,\rho\,]),
\\
\label{cfunctional} 
c(\rv_1,\rv_2)&=c(\rv_1,\rv_2;[\,\rho\,]),
\end{align}
where the square brackets indicate a functional dependence. 
Given the functional \eqref{cfunctional} the LMBW equation \eqref{LMBW} can be solved self-consistently for the one-body 
density, for any external potential.

\subsection{Percus functional}

DFT is an exact formalism for the study of inhomogeneous many-body systems \cite{evans79,evans92}.  
The main quantity of interest, the grand potential functional, 
is given by
\begin{align}\label{grand}
\Omega[\,\rho\,] = F^{ \,\text{id}}[\,\rho\,] + F^{ \,\text{exc}}[\,\rho\,] 
- \int \!d\rv \big( \mu - V_{\text{ext}}(\rv) \big)\rho(\rv), 
\end{align}
where $\mu$ is the chemical potential.
While the Helmholtz free energy of the ideal gas is known exactly,  the excess part, $F^{\text{exc}}[\,\rho\,]$, encodes the interparticle 
interactions and usually has to be approximated.
The Euler-Lagrange (EL) equation 
for the density is generated by the following variational condition
\begin{align}
\label{EQomegaMinimial}
\frac{\delta  \Omega[\rho\,]}{\delta \rho(\rv)}=0
\hspace*{0.2cm}
\rightarrow
\hspace*{0.2cm}
\rho(\rv)=e^{ -\beta\left(V_{\text{ext}}(\rv) - \mu - k_BTc^{(1)}(\rv)\right)},
\end{align}
where the one-body direct correlation function is defined according to 
$c^{(1)}(\rv)\equiv-\delta \beta F^{\text{exc}}/\delta\rho(\rv)$. 
Given an explicit form for $F^{\text{exc}}$, equation 
\eqref{EQomegaMinimial} can be solved to determine the equilibrium density corresponding to the chosen external field. 
Substitution of the equilibrium density back into the grand potential functional \eqref{grand} then yields the equilibrium grand potential of the inhomogeneous system.

A system of hard rods in 1D (the inhomogeneous Tonks gas) is one of the rare cases for which the excess Helmholtz free energy functional can be determined 
exactly \cite{percus_hard_rods,evans92}. 
It is given by
\begin{equation}\label{percus_functional}
    F^{\text{exc}}[\,\rho\,]=-k_BT\int_{-\infty}^{+\infty}\!dx \,  n_0(x)\ln (1-n_1(x)) , 
\end{equation}
where the weighted densities $n_0$ and $n_1$ are generated 
from the convolution integrals
\begin{equation}
    \begin{aligned}
    &n_{\alpha}(x) = \int_{-\infty}^{+\infty}\!dx' \rho(x')
    w_{\alpha}(|x-x'|),
    \end{aligned}
\end{equation}
which employ geometrical weight functions given by
\begin{align}\label{weight_ends}
    w_0(x) &= \frac{1}{2}\Big( 
    \delta(x-R)+\delta(x+R)\Big),
\\
\label{weight_volume}
w_1(x)&=\begin{cases}
    1, & -R < x < R,\\
    0, & \text{otherwise},
  \end{cases}
\end{align}
where $R$ is the particle radius.
The Percus functional \eqref{percus_functional} expresses the excess Helmholtz free 
energy as an integral over a free energy density, which is a 
local function of two weighted densities. These are constructed using weight functions reflecting the `surface' \eqref{weight_ends} and `volume' \eqref{weight_volume} 
of the rods. 
This structure served as inspiration for Rosenfeld's famous 3D hard-sphere functional \cite{Rosenfeld89} which subsequently led to the development of a new class of approximations, generically referred to as fundamental measure theory (FMT)     
\cite{Tarazona0Dreview,RothReview}.

Solution of the EL equation \eqref{EQomegaMinimial} using the functional \eqref{percus_functional} yields the equilibrium one-body density for any given external potential. 
Within this scheme, information about the two-body correlations in the inhomogeneous fluid can be obtained from a second functional derivative of 
the excess Helmholtz free energy 
\begin{align}\label{2derivative}
c(\rv_1,\rv_2)=-\frac{\delta^2 
\beta F^{ \,\text{exc}}[\,\rho\,]} {\delta\rho(\rv_1)\delta\rho(\rv_2)}.
\end{align}
This determines the two-body direct correlation function as 
a functional of the one-body density. 
Evaluation of \eqref{2derivative} using the equilibrium one-body density and then substituting the result into the inhomogeneous OZ equation 
\eqref{Inhomogeneous OZ equation} yields the total correlation function. 
In this way the Percus functional \eqref{percus_functional} provides 
exact results for both the one-body density and inhomogeneous two-body correlation functions of 1D hard rods. 
In the absence of an external field (setting aside any of the subtleties related to phase coexistence) the one-body density reduces to a constant, $\rho(\rv)\!\rightarrow \rho_b$, and the two-body direct correlation function \eqref{2derivative} becomes translationally and rotationally invariant, $c(\rv_1,\rv_2)\!\rightarrow\! c(r_{12})$.

\subsection{Dimensional crossover}

\begin{figure}[!t]
\includegraphics[width=\linewidth]{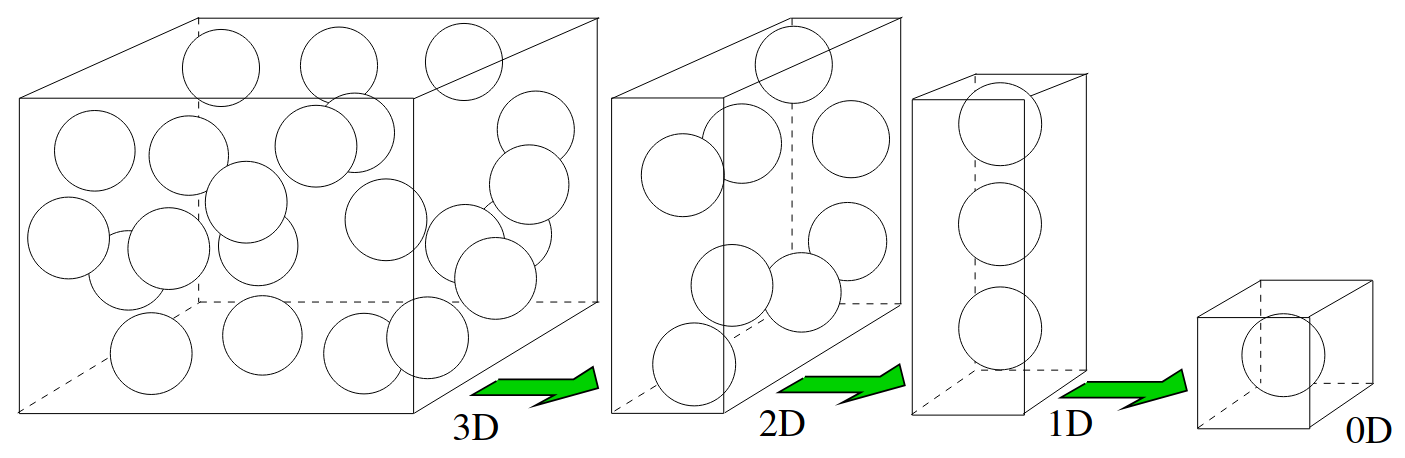}
\caption{\textbf{Dimensional crossover imposed by the sequential application of confining potentials.} For example, trapping  a 3D system of hard spheres between parallel hard walls with a separation $L\!=\!d$ recovers a 2D bulk system of hard disks. Introducing a further pair of confining walls then reduces the system to 1D hard rods on a line. Finally, the 0D limit is obtained when the external boundaries create a cavity which can admit at most one particle.}
\label{fig crossover 3 to 0}
\end{figure}

When a system is confined between two closely spaced walls (see Fig.~\ref{fig crossover 3 to 0}), 
it forces the particles to move only in the 2D space available to them. The 3D density profile then reduces to the form
\begin{equation}\label{reduction1}
\rho_{\,\text{3D}}(x,y,z)\!\rightarrow\!\rho_{\,\text{2D}}(x,y)\delta(z), 
\end{equation}
where we have chosen to restrict motion in the $z$-direction.
Further confining the particles to move on a line in the $x$-direction yields
\begin{equation}\label{reduction2}
\rho_{\,\text{3D}}(x,y,z)\!\rightarrow\!\rho_{\,\text{1D}}(x)\delta(y)\delta(z). 
\end{equation}
The ultimate limit of this process of dimensional reduction 
is a zero-dimensional state which, assuming the interaction potential does not allow full particle overlap, corresponds to a cavity which can hold either zero or one particle. 
In this case we would have 
\begin{equation}\label{reduction3}
\rho_{\,\text{3D}}(x,y,z)\!\rightarrow\!\eta\,\delta(x)\delta(y)\delta(z),
\end{equation}
where $\eta$ is the zero-dimensional packing fraction \cite{TarazonaRosenfeld_0D}. 
Correct handling of the 0D limit was found to be very important to obtain physical results from standard applications of DFT to bulk freezing, for which the one-body density is modelled as an array of localized density peaks centered on lattice sites \cite{evans92,Tarazona_Freezing}. 
If approximations to the excess free energy functional in \eqref{grand} are not very carefully constructed, then these can diverge and yield unphysical results 
as the dimensionality is reduced.
Indeed, earlier attempts to treat crystalline states encountered severe 
divergences in the 0D limit \eqref{reduction3}.
In contrast, the \textit{exact} excess Helmholtz free energy functional of a 3D system, $F^{\text{\,exc}}_{3D}$, would handle correctly any of the density profiles \eqref{reduction1}-\eqref{reduction3}. 
For example, taking the profile \eqref{reduction1} as input would yield  
\begin{align}\label{crossover1}
F^{\text{\,exc}}_{3D}[\,\rho_{\,\text{2D}}(x,y)\delta(z)\,]
=
F^{\text{\,exc}}_{2D}[\,\rho_{\,\text{2D}}(x,y)\,],  
\end{align}
where $F^{\text{\,exc}}_{2D}$ is the exact 2D functional. 
This would then further 
reduce to the exact 1D functional, according to
\begin{align}\label{crossover2}
F^{\text{\,exc}}_{2D}[\,\rho_{\,\text{1D}}(x)\delta(y)\,]
=
F^{\text{\,exc}}_{1D}[\,\rho_{\,\text{1D}}(x)\,]. 
\end{align}
Finally, in the 0D limit we would recover the exact free energy of a single occupancy cavity, namely
\begin{align}\label{crossover3}
F^{\text{\,exc}}_{1D}[\,\delta(x)\,]
=
F^{\text{\,exc}}_{0D}(\eta).  
\end{align}
For the case of hard rods it is a straightforward exercise to show that the exact Percus functional \eqref{percus_functional} 
satisfies the condition \eqref{crossover3}. 

Rosenfeld and Tarazona exploited the idea of dimensional 
crossover to help develop new forms for the excess Helmholtz free energy functional within FMT \cite{RosenfeldTarazonaQuasi,TarazonaRosenfeld_0D,Tarazona_Freezing}. 
One of the challenges with this approach is to obtain functionals 
which possess the desired dimensional crossover properties, while still remaining accurate in bulk. 
The 3D hard-sphere FMT functional developed in \cite{Tarazona_Freezing} reduces correctly to the exact expressions in both the 0D and 1D limits and recovers the 
PY equation of state and PY pair direct correlation function in the bulk. 
(The quality of the two-dimensional functional which emerges from this approximation when taking the limit from 3D to 2D remains 
unclear.)
In the same study Rosenfeld and Tarazona also made steps towards an FMT functional for the 2D 
hard-disk system \cite{TarazonaRosenfeld_0D}. However, 
this was not expressed in an implementable closed form, 
since the geometry of the 2D system neccessitates an excess Helmholtz free energy dependent on an infinite number of weighted densities. 
A good compromise was reached by Roth, Mecke and Oettel, who employed a finite set of weighted densities (truncating the infinite series) and then enforced the PY equation of state 
to fix free parameters \cite{OettelDisk}. 
This 2D functional recovers both the correct 0D limit and the 
PY equation of state for bulk hard disks but, due to the neglect of higher-order weighted densities, does not recover exactly the Percus functional in the 1D limit.   
More recently, Lutsko has used the 0D limit to construct an FMT approximation for 3D hard spheres which performs well at higher packing but, as a neccessary compromise, does not satisfy the exactly known low-density limit \cite{lutsko_0D}. 
One conclusion that can be drawn from the aforementioned works on 
FMT is that it is difficult to construct a free energy functional exhibiting consistent dimensional crossover, while still maintaining a uniformly good account of the bulk fluid. 
Approximations have to be implemented on a case-by-case basis and each functional must be custom-made according to the pair interaction potential and the dimensionality of the native bulk system. 
Moreover, the 2D limit of a 3D FMT functional will 
in general not coincide with the functional obtained by applying FMT specifically to the 2D system.

In subsection \ref{crossoverPY} we will show how the inhomogeneous integral equation method handles dimensional crossover in a very natural way and completely avoids the difficulties encountered 
when seeking to find explicit forms for the excess Helmholtz free energy. Our approach is free from constraint-induced divergences and easily recovers sensible bulk limits.

\subsection{Hard disks under quasi one-dimensional confinement: An exact solution}\label{Q1D exact}

\begin{figure}[!t]
\vspace*{0.3cm}
1. 2D slit, with $L \gg d$:
\begin{center}
\begin{tikzpicture}
\coordinate (origine) at (0,0);
\coordinate (particle 1) at (-1.2,0.8);
\coordinate (particle 2) at (-2.45,-0.4);
\coordinate (particle 3) at (-0.9,-0.9);
\coordinate (particle 4) at (0.6,-0.1);
\coordinate (particle 5) at (1.85,0.7);
\coordinate (particle 6) at (2,-0.7);
\coordinate (upper wall left end) at (-3.25,1.5);
\coordinate (upper wall right end) at (2.8,1.5);
\coordinate (upper wall size) at (3.25+2.8,0.75);
\coordinate (lower wall left end) at (-3.25,-1.5);
\coordinate (lower wall right end) at (2.8,-1.5);
\coordinate (lower wall size) at (3.25+2.8,-0.75);
\draw[-, >=latex, line width=0.8] (upper wall left end) -- (upper wall right end);
\draw[-, >=latex, line width=0.8] (lower wall left end) -- (lower wall right end);
\fill[pattern=north west lines] (upper wall left end) rectangle ++(upper wall size);
\fill[pattern=north west lines] (lower wall left end) rectangle ++(lower wall size);
\draw[line width=0.7] (particle 1) circle (15pt);
\draw[line width=0.7] (particle 2) circle (15pt);
\draw[line width=0.7] (particle 3) circle (15pt);
\draw[line width=0.7] (particle 4) circle (15pt);
\draw[line width=0.7] (particle 5) circle (15pt);
\draw[line width=0.7] (particle 6) circle (15pt);
\draw[<->, >=latex, line width=0.8, gray] (-0.9cm-11pt,-0.9cm-11pt) -- (-0.9cm+11pt,-0.9cm+11pt);
\draw[gray] (-0.8,-0.9) node[above left] {\normalsize $d$};
\draw[<->, >=latex, line width=0.8, gray] (upper wall left end) -- (lower wall left end);
\draw[gray] (-3.25,1) node[right] {\normalsize $L$};
\end{tikzpicture}
\end{center}
\vspace*{0.3cm}
2. Quasi-1D slit, with $d<L<2\,d$:
\begin{center}
\begin{tikzpicture}
\coordinate (origine) at (0,0);
\coordinate (particle 1) at (-2.45,0.15);
\coordinate (particle 2) at (-0.9,-0.21);
\coordinate (particle 3) at (0.62,0.17);
\coordinate (particle 4) at (2.1,0.11);
\coordinate (upper wall left end) at (-3.25,2-1.2);
\coordinate (upper wall right end) at (2.8,2-1.2);
\coordinate (upper wall size) at (3.25+2.8,0.75);
\coordinate (lower wall left end) at (-3.25,-2+1.2);
\coordinate (lower wall right end) at (2.8,-2+1.2);
\coordinate (lower wall size) at (3.25+2.8,-0.75);
\draw[-, >=latex, line width=0.8] (upper wall left end) -- (upper wall right end);
\draw[-, >=latex, line width=0.8] (lower wall left end) -- (lower wall right end);
\fill[pattern=north west lines] (upper wall left end) rectangle ++(upper wall size);
\fill[pattern=north west lines] (lower wall left end) rectangle ++(lower wall size);
\draw[line width=0.7] (particle 1) circle (15pt);
\draw[line width=0.7] (particle 2) circle (15pt);
\draw[line width=0.7] (particle 3) circle (15pt);
\draw[line width=0.7] (particle 4) circle (15pt);
\end{tikzpicture}
\end{center}
\vspace*{0.3cm}
3. 1D confinement, with $L\rightarrow d$:
\begin{center}
\begin{tikzpicture}
\coordinate (origine) at (0,0);
\coordinate (particle 1) at (-2.45,0.02);
\coordinate (particle 2) at (-0.9,0);
\coordinate (particle 3) at (0.62,0.01);
\coordinate (particle 4) at (2.1,-0.01);
\coordinate (upper wall left end) at (-3.25,2-1.4);
\coordinate (upper wall right end) at (2.8,2-1.4);
\coordinate (upper wall size) at (3.25+2.8,0.75);
\coordinate (lower wall left end) at (-3.25,-2+1.4);
\coordinate (lower wall right end) at (2.8,-2+1.4);
\coordinate (lower wall size) at (3.25+2.8,-0.75);
\draw[-, >=latex, line width=0.8] (upper wall left end) -- (upper wall right end);
\draw[-, >=latex, line width=0.8] (lower wall left end) -- (lower wall right end);
\fill[pattern=north west lines] (upper wall left end) rectangle ++(upper wall size);
\fill[pattern=north west lines] (lower wall left end) rectangle ++(lower wall size);
\draw[line width=0.7] (particle 1) circle (15pt);
\draw[line width=0.7] (particle 2) circle (15pt);
\draw[line width=0.7] (particle 3) circle (15pt);
\draw[line width=0.7] (particle 4) circle (15pt);
\end{tikzpicture}
\end{center}
\caption{\textbf{Sketch of dimensional crossover from 2D to 1D.}
The first panel shows a system of hard-disk particles in a two-dimensional slit, with two hard walls separated by a distance $L$ much larger than one particle diameter, $d$.
The second panel shows a slit with a wall separation between one and two particle diameters.
In this case the channel is too narrow to allow particles to move past each other. 
We refer to this as the quasi-1D regime.
The third panel shows a slit with a wall separation very close to one particle diameter, for which the system reduces to the one-dimensional bulk system of hard rods on a line.
}
\label{fig geometry dim reduction}
\end{figure}
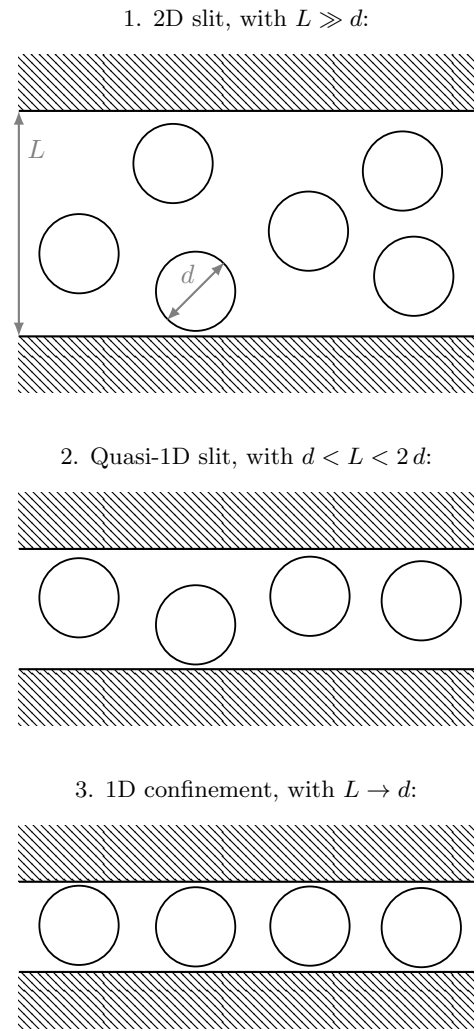

We will consider 2D hard disks, for which the interparticle interaction potential is given by
\begin{align}\label{disk potential}
\phi(r_{12}) = 
\begin{cases}
      \;\;0, & r_{12}\ge d, \vspace*{0.1cm}\\
      \;\infty, & r_{12} < d, 
\end{cases}
\end{align}
where $r_{12}\!=\!(\, (x_1-x_2)^2 + (z_1-z_2)^2 \,)^{\frac{1}{2}}$. 
The external potential representing two parallel hard walls 
separated by a distance $L$ in the $z$-direction is given by
\begin{align}\label{wall potential}
V_{\text{ext}}(z) = 
\begin{cases}
      \;\;0,  &-\frac{L}{2}+R < z < \frac{L}{2}-R, \vspace*{0.1cm}\\
      \;\infty, & \text{otherwise}.
\end{cases}
\end{align}
The equilibrium one-body density exhibits the same symmetry 
as the external potential and is thus translationally invariant 
in the $x$-direction, $\rho(\rv)\!=\!\rho(z)$. 
Since we will be considering hard disks confined to a 2D channel it is convenient to characterize the degree of packing 
in the system using the average number of particles per unit channel length, $\langle N\rangle$. 
As we set $d\!=\!1$, $\langle N\rangle$ is given by
\begin{equation}
\langle N\rangle=
\int_{-\infty}^{+\infty} \!\!dz\, 
\rho(z).
\end{equation}
Due to translational invariance the arguments of the two-body total and direct correlation functions can be simplified 
according to 
\begin{align}
h(\rv_1,\rv_2)&\rightarrow h(z_1,z_2,x_{12}),
\label{translational invariant h}
\\
c(\rv_1,\rv_2)&\rightarrow c(z_1,z_2,x_{12}),
\label{translational invariant c}
\end{align}
where $x_{12}\!=\!|x_1-x_2|$ is the longitudinal distance 
between the two points of interest. 
When presenting results for the inhomogeneous correlations in the channel we will focus on the 
pair distribution function, $g$. This physically intuitive quantity is related to the 
total correlation function by a trivial shift,
\begin{equation}
g(z_1,z_2,x_{12}) = h(z_1,z_2,x_{12}) + 1.
\end{equation}
For channel widths in the range $1<L<2$ the system is in the quasi-1D regime. 

In Ref.~\cite{Q1D_santos_structure}, Montero and Santos presented a novel approach to exact solution of the quasi-1D hard-disk system by mapping it onto a multicomponent system of one-dimensional hard rods. 
The length of the effective 1D rods represents the difference 
in the transverse positions of pairs of adjacent disks. 
Since this mapping is exact it becomes possible to retrieve the pair distribution function of the quasi-1D system using 
statistical-mechanical methods developed for treating hard-rod mixtures.
Although the solutions presented in \cite{Q1D_santos_structure} are exact, the expressions are rather complicated and their 
numerical implementation is still not entirely straightforward. We could thus profit from the freely available code 
provided by the authors of \cite{Q1D_santos_structure} to generate reference data against which to compare our numerical solution of the inhomogeneous PY integral equation. 
The expressions that generate the exact quasi-1D pair distribution functions are entirely analytical and do not require any iterative numerical solution. 
This enables $g$ to be calculated rapidly for any value of 
$\langle N\rangle$ below close packing without stability 
or convergence issues. 
Moreover, the results are independent of both the grid spacing 
and the length of the computational domain along the channel, which is not the case when performing numerical solution of 
the inhomogeneous PY theory.
Nevertheless, the latter offers much more flexibility, since it 
remains valid for $L>2$ (i.e.~when the system becomes truly 
2D) while the hard-rod mapping breaks down. 
We note that Montero and Santos have very recently given 
an exact solution for the quasi-1D system consisting of 3D 
hard spheres confined within a narrow tube \cite{Santos_tube}.

\section{Results}

\subsection{Dimensional reduction properties of the PY closure}

\label{crossoverPY}

\subsubsection{Numerical results}

\begin{figure}[t!]
\includegraphics[width=\linewidth]{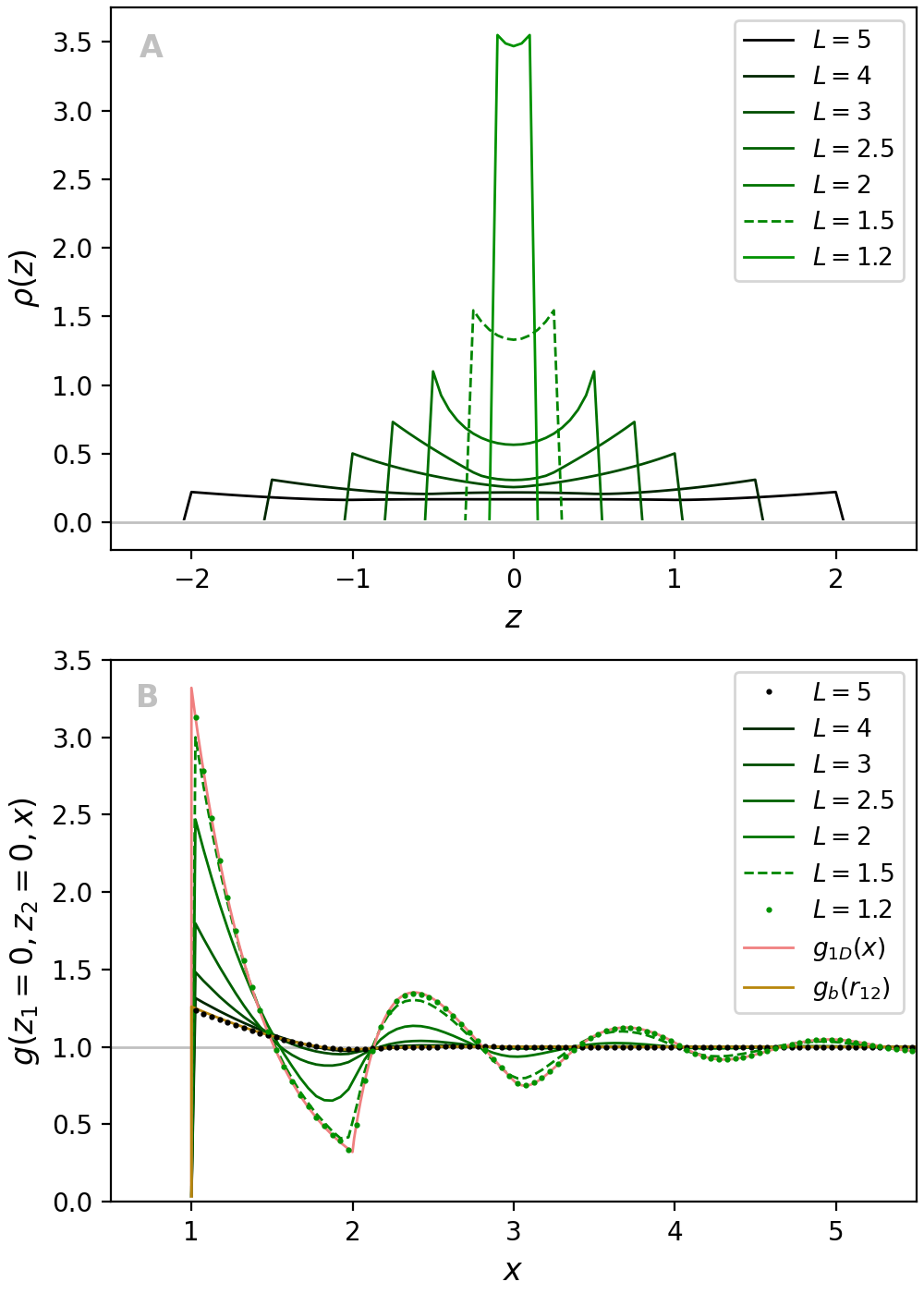}
\caption{\textbf{Dimensional crossover from hard disks to hard rods.}
Panel A shows, in shades of green, the one-body density profiles for a system of hard-disk particles between two hard walls, as the slit width $L$ is reduced from $5$ to $1.2$ (for particle diameter $d$ set to unity). For all profiles the average number of particles per unit channel length is kept constant at the value $\avN\!=\!0.7$. 
Panel B shows, in the same shades of green, the evolution of the pair distribution function along a line in the center of the channel, namely 
$g(z_1\!=\!0,z_2\!=\!0,x)$, as the slit width $L$ is reduced toward unity.
The bulk two-dimensional hard-disk radial distribution function, 
$g_{\text{b}}(r_{12})$, evaluated at $\rho_{\text{b}}\!=\!\avN/(L-1)$ is shown as a solid light brown line and the exact radial distribution function of hard rods along the $x$-axis, 
$g_{\text{1D}}(x)$, is shown as a solid pink line.
}
\label{fig results dim reduction}
\end{figure}

Starting from a two-dimensional slit, with separation length $L\!=\!5$ between the hard walls (given in units of particle diameters), we aim to observe how the inhomogeneous two-body correlation function, $g$, of a system of hard-disk particles behaves when $L$ is reduced to approach a value of one and assess if the dimensional crossover to a bulk 1D system of hard rods (as described in Fig.~\ref{fig geometry dim reduction}) is exactly recovered. Results for a fixed average number of particles per unit length $\langle N \rangle \!=\! 0.7$, obtained from numerical solution of equations 
\eqref{Inhomogeneous OZ equation}, \eqref{LMBW} and \eqref{HD PY closure}, are shown in Fig.~\ref{fig results dim reduction}.

In panel A we show the obtained one-body density profiles for values of $L$ starting at $5$ and decreasing to $1.2$ (dark to light shades of green). As $L$ is slightly modified from $1.5$ to $1.2$ the density profile is hugely impacted and becomes much more peaked in the center of the available slit. This is consistant with the fact that for $L\!\rightarrow\!1$ the system becomes one-dimensional and $\rho(z)\!\rightarrow\!\avN\,\delta(z)$.

In panel B we show the inhomogeneous two-body correlation function along a line in the center of the channel, $g(z_1\!=\!0,z_2\!=\!0,x)$, for the same values of $L$ (and in the same shades of green).
We used dotted lines for both the biggest and the narrower slits (darkest and lightest shades, respectively)  to compare them with their expected limits.
For the widest channel width shown, $L\!=\!5$, we recover the expected bulk two-dimensional hard-disk radial distribution function, 
$g_{\text{b}}(r_{12})$, evaluated at $\rho_{\text{b}}\!=\!\avN/(L-1)$ and shown as a solid light brown line.
As the channel width is reduced below $L\!=\!2$ the system becomes quasi one-dimensional. 
We see that already at $L\!=\!1.2$ we recover very accurately the exact radial distribution function of hard rods along the $x$-axis (Tonks gas), $g_{\text{1D}}(x)$, shown as a solid pink line.

\subsubsection{Dimensional reduction of the OZ equation}

The preceding numerical results demonstrate that as $L$ is reduced the pair 
distribution function along the center of the channel converges rapidly to the exact 1D radial distribution function of 
hard rods. However, for values of $L$ very close to unity the one-body density approaches a 
delta-function 
and accurate numerical resolution becomes challenging. 
We will thus consider analytically the limit 
$L\rightarrow 1$ to elucidate how the 2D inhomogeneous OZ equation indeed reduces in a very natural way to its exact one-dimensional counterpart.
In this limit, we have 
\begin{equation}\label{confine to channel}
\rho(\rv)
=\rho(x,z)
\rightarrow\langle N\rangle\,\delta(z),
\end{equation}
where we recall that $\langle N\rangle$ is the average number of particles per unit channel length.
Using the translationally invariant forms \eqref{translational invariant h} 
and \eqref{translational invariant c} for the total and direct correlation functions, respectively, the 
inhomogeneous OZ equation \eqref{Inhomogeneous OZ equation} can be expressed as follows
\begin{align}\label{OZ2D}
&h(z_1,z_2,x_{12}) = c(z_1,z_2,x_{12})
\\ 
& + \langle N\rangle\!\!\int \!dx_3\!\int \!dz_3\; h(z_1,z_3,x_{13})\,
\delta(z_3) 
\, c(z_2,z_3,x_{32}).
\notag
\end{align}
In the limit $L\rightarrow\!1$ the only nonzero correlations are for points on the center line, $z_1\!=\!z_2\!=\!0$, where the one-body density is not vanishing. 
Substitution of \eqref{confine to channel} into \eqref{OZ2D} eliminates the integral over $z_3$ to yield
\begin{align}\label{OZ2Dto1D}
h(0,0,x_{12}) &= c(0,0,x_{12}) 
\\
&+ 
\langle N\rangle\!\! \int\! dx_3\; h(0,0,x_{13})\,c(0,0,x_{32}),
\notag
\end{align}
which is the bulk OZ equation \eqref{bulk OZ equation} in 1D, 
albeit expressed in a slightly unfamiliar way.  
This can be made clearer by changing the notation to 
\begin{align}
\langle N\rangle &= \rho^{\text{1D}}_{\text{b}},
\\
h(0,0,x_{12}) &= h^{\text{1D}}(x_{12}),
\\
c(0,0,x_{12}) &= c^{\text{1D}}(x_{12}),
\end{align}
which puts \eqref{OZ2Dto1D} into the following standard form
\begin{equation}
\!\!h^{\text{1D}}(x_{12}) = c^{\text{1D}}(x_{12}) 
\,+\, 
\rho^{\text{1D}}_{\text{b}}\!\! \int\! dx_3\; h^{\text{1D}}(x_{13})\,c^{\text{1D}}(x_{32}).
\end{equation}
The two-body correlation functions along the central line are subject to the PY constraints
\begin{align}
h(0,0,x_{12})&=-1,\quad r_{12}<1, \notag\\
c(0,0,x_{12})&=0,\quad\;\;\; r_{12}>1, 
\end{align}
which is evidently the same as 
\begin{align}
h^{\text{1D}}(x_{12})&=-1,\quad x_{12}<1, \notag\\
c^{\text{1D}}(x_{12})&=0,\quad\;\;\; x_{12}>1.
\end{align}
It is thus apparent that the PY closure \eqref{HD PY closure} crosses over naturally to 1D.
Finally, when equation \eqref{confine to channel} is substituted into the LMBW equation \eqref{LMBW} we directly obtain the result 
that the gradient of the one-body density is zero along the channel, consistent with a constant 1D longitudinal density profile. 
As mentioned previously, the PY closure is exact for 1D hard rods.

\begin{figure}[!t]
\vspace*{0.3cm}
\begin{center}
\begin{tikzpicture}
\coordinate (origine) at (0,0);
\coordinate (particle 1) at (-2.45,0.26);
\coordinate (particle 2) at (-1.55,-0.26);
\coordinate (particle 3) at (-0.65,0.26);
\coordinate (particle 4) at (0.25,-0.26);
\coordinate (particle 5) at (1.15,0.26);
\coordinate (particle 6) at (2.05,-0.26);
\coordinate (upper wall left end) at (-3.25,2-1.2);
\coordinate (upper wall right end) at (2.8,2-1.2);
\coordinate (upper wall size) at (3.25+2.8,0.75);
\coordinate (lower wall left end) at (-3.25,-2+1.2);
\coordinate (lower wall right end) at (2.8,-2+1.2);
\coordinate (lower wall size) at (3.25+2.8,-0.75);
\draw[-, >=latex, line width=0.8] (upper wall left end) -- (upper wall right end);
\draw[-, >=latex, line width=0.8] (lower wall left end) -- (lower wall right end);
\fill[pattern=north west lines] (upper wall left end) rectangle ++(upper wall size);
\fill[pattern=north west lines] (lower wall left end) rectangle ++(lower wall size);
\draw[line width=0.7] (particle 1) circle (15pt);
\draw[line width=0.7] (particle 2) circle (15pt);
\draw[line width=0.7] (particle 3) circle (15pt);
\draw[line width=0.7] (particle 4) circle (15pt);
\draw[line width=0.7] (particle 5) circle (15pt);
\draw[line width=0.7] (particle 6) circle (15pt);
\draw[->, >=latex, line width=1.1, chromeyellow] (-3.25,0.30) -- (2.8,0.26);
\draw[chromeyellow] (2.25,0.25) node[above] {\small path $2$};
\draw[->, >=latex, line width=1.1, bostonuniversityred] (-3.25,0) -- (2.8,0);
\draw[bostonuniversityred] (-2.8,-0.2) node[below] {\small path $1$};
\end{tikzpicture}
\end{center}
\caption{\textbf{Densely packed hard disks in a quasi-1D slit.}
For a wall separation $L\!=\!1.5$ we show the maximally packed zigzag state, for which the number of particles per unit channel length, $\langle N\rangle$, is equal to $2/\sqrt{3}$.
Red and yellow arrows indicate the paths along which we will show results for the inhomogeneous 
pair distribution function in Fig.~\ref{fig results quasi 1 dim g}. 
Path 1 follows the center of the channel, $z\!=\!0$, whereas path 2 is located at 
$z\!=\!0.25$, corresponding to disks in contact with the upper wall.
}
\label{fig geometry quasi 1 dim 1p5}
\end{figure}
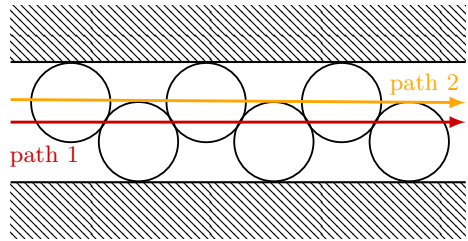

We would like to emphasise here that the longitudinal one-body density profile of an infinitely long 1D channel is just a constant, equal to the average number of particles per unit channel length. This is a simple consequence of translational invariance. 
If the rods were very tightly packed (i.e.~$\langle N\rangle\!\rightarrow\!1$) then one may be tempted to think 
that the longitudinal density profile should approach a `Dirac comb', namely a sum of delta-functions each separated from its neighbour by a rod length. 
However, the Boltzmann statistics underlying our formalism include averaging over a global translation, which ensures a density profile consistent with the symmetries imposed by the external potential. 
Longitudinal order \textit{within} the translationally 
invariant 1D fluid is revealed by the two-body correlation functions, not the longitudinal one-body density profile.

\subsection{Quasi one-dimensional confinement}

In the quasi-1D regime the maximally packed state is commonly referred to as the `zigzag state'.
(For illustration we show, in Fig.~\ref{fig geometry quasi 1 dim 1p5}, a sketch for the case $L\!=\!1.5$.)
Investigating the development of such a zigzag state when the packing is increased is of fundamental 
interest, since it mimics in a simplified way the crystallization and ordering processes observed 
at substrates in larger, more realistic systems \cite{Dijkstra}.  
At high packing, approaching the zigzag state, the system possesses both a nontrivial density 
profile, $\rho(z)$, and inhomogeneous pair correlation functions which exhibit the emergence of long-range ordering parallel to the walls. 
Since these two physical effects are generic to all substrate ordering phenomena the present, exactly soluble, quasi-1D slit model provides an excellent opportunity to test the capabilities of the inhomogeneous PY theory. 
If the approximate theory can give a good account of the exact quasi-1D solution, then one can proceed with 
confidence to address more demanding problems of substrate ordering for which exact solutions are 
no longer available.

\subsubsection{Numerical results}

We consider a slit of fixed width $L\!=\!1.5$ to study how the density profile and pair distribution 
function predicted by the PY theory evolve as the packing is increased. 
Our numerical results are benchmarked against the exact solution of Montero and Santos \cite{Q1D_santos_structure}.

\begin{figure}[t!]
\includegraphics[width=\linewidth]{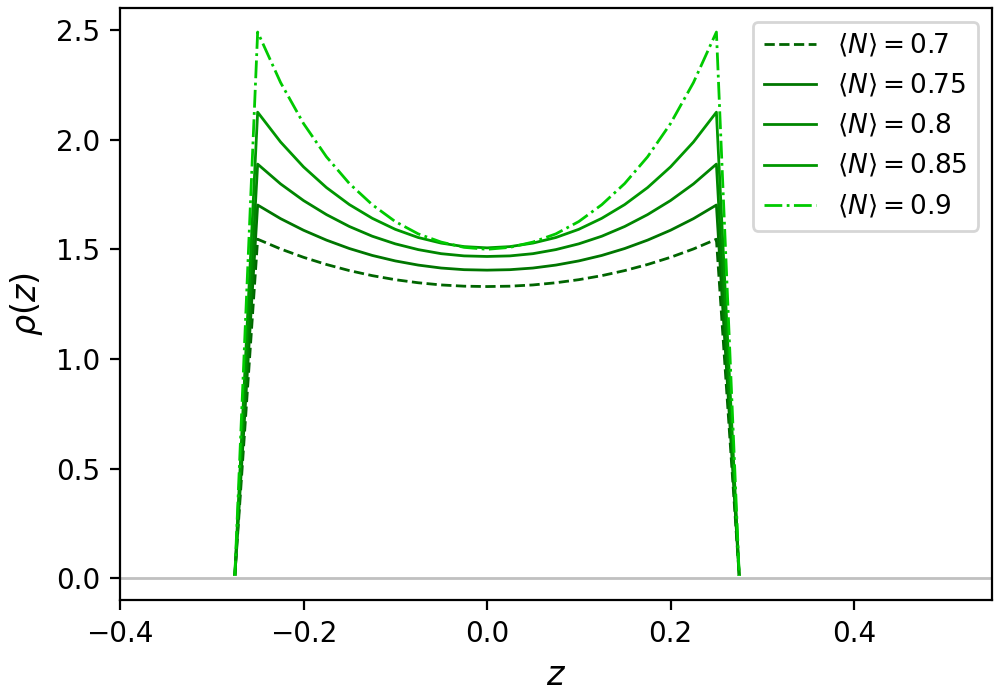}
\caption{\textbf{Density profiles in the quasi one-dimensional regime.} We show one-body densities for hard disks trapped in a channel with fixed separation length $L\!=\!1.5$ and for increasing packing. Starting from an average number of particles per unit length $\langle N \rangle \!=\! 0.7$ we recover the exact same dashed green curve than already shown previously in panel A of Fig.~\ref{fig results dim reduction}. The density profiles for $\langle N \rangle \!=\! 0.75, 0.8$ and $0.85$ are shown as solid lines in different shades of green.
The density profile for $\langle N \rangle \!=\! 0.9$, shown as a dotted-dashed light green curve,
exhibits the most pronounced contact peaks. However, its value at the center of the channel, for $z\!=\!0$, is lower than that for $\langle N \rangle \!=\! 0.85$.  
}
\label{fig results quasi 1 dim rho}
\end{figure}
\begin{figure}[t!]
\includegraphics[width=\linewidth]{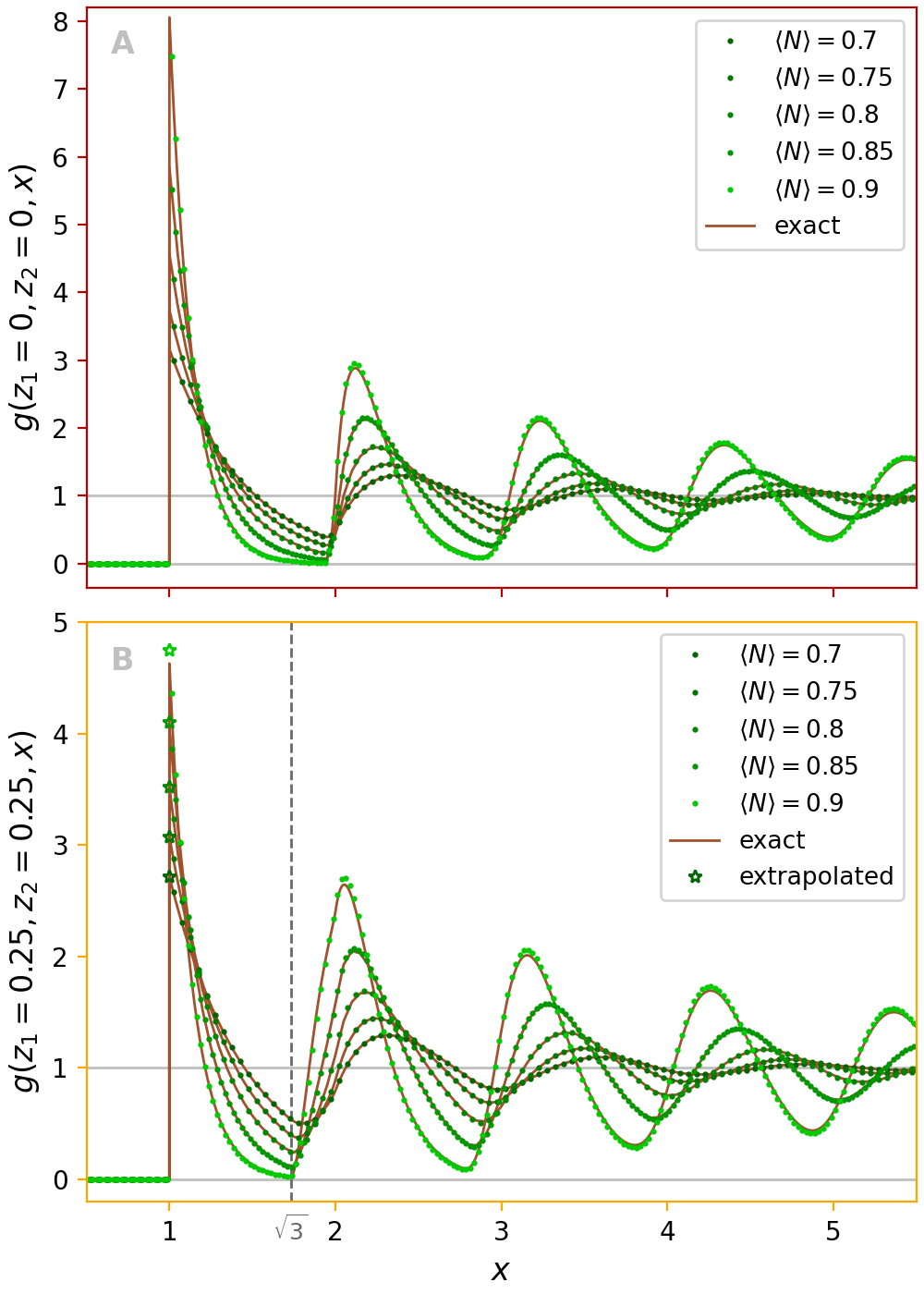}
\caption{\textbf{Inhomogeneous pair distribution function in the quasi-1D regime.}
We show, as green dotted lines, the inhomogeneous two-body correlation function $g$, plotted with respect to variable $x$ for $z_1\!=\!z_2$,  corresponding to the density profiles of Fig.~\ref{fig results quasi 1 dim rho}. The exact results for $g$ obtained in \cite{Q1D_santos_structure} are displayed as solid brown lines. Panel A (framed in red) and panel B (in yellow) show the case for $z_1\!=\!z_2\!=\!0$ (path 1 in Fig.~\ref{fig geometry quasi 1 dim 1p5}) and $z_1\!=\!z_2\!=\!0.25$ (path 2 in Fig.~\ref{fig geometry quasi 1 dim 1p5}), respectively. The dashed grey vertical line at $x\!=\!\sqrt{3}$ in panel B shows the distance of closest approach for next-nearest neighbours in the zigzag phase along path 2.
In both panels, for $\langle N \rangle\!=\!0.7$ to $0.9$, the green dotted lines agree almost perfectly with the exact results. For panel B we give the extrapolated contact values at $x\!=\!1$ as green stars.
}
\label{fig results quasi 1 dim g}
\end{figure}

In Fig.~\ref{fig results quasi 1 dim rho} we show the one-body density profiles for values of 
$\langle N\rangle$ between $0.7$ and $0.9$ obtained from numerical solution of equations 
\eqref{Inhomogeneous OZ equation}, \eqref{LMBW} and \eqref{HD PY closure}.
For the lowest packing, $\langle N\rangle\!=\!0.7$, we reproduce the dashed green curve 
already shown in panel A of Fig.~\ref{fig results dim reduction}. 
Density profiles for $\langle N \rangle \!=\! 0.75, 0.8$ and $0.85$ are shown as solid lines in different shades of green. 
All of these density profiles are similar in form, reflecting the accumulation of particles at the walls and depletion from the center of the channel. 
As $\langle N\rangle$ is increased from $0.7$ to $0.85$ both the density in the channel center, $z\!=\!0$, and at contact, $z\!=\!\pm 0.25$, increase.  
This trend can be seen to change for $\langle N \rangle \!=\! 0.9$, shown as a dotted-dashed light green curve. 
Indeed, although the height of the two contact peaks continues to increase in going from 
$\langle N \rangle \!=\! 0.85$ to $0.9$, the value of the density at the center has decreased.
This noticible change of behaviour signifies the onset of the zigzag state, as will be discussed in-depth below.
Note that, in the maximally packed limit, $\langle N\rangle\!=\!2/\sqrt{3}\!\approx\!1.1547$, the density profile will become a sum of two Dirac delta-functions located at contact, which cannot be resolved numerically.

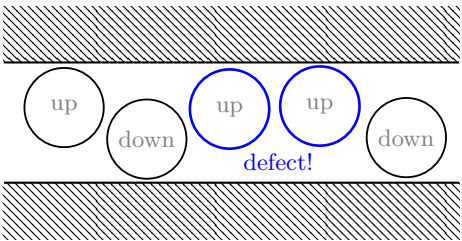
\begin{figure}[!t]
\begin{center}
\begin{tikzpicture}
\coordinate (origine) at (0,0);
\coordinate (particle 1) at (-2.45,0.2);
\coordinate (particle 2) at (-1.35,-0.22);
\coordinate (particle 3) at (-0.25,0.18);
\coordinate (particle 4) at (0.95,0.22);
\coordinate (particle 5) at (2.1,-0.2);
\coordinate (upper wall left end) at (-3.25,2-1.2);
\coordinate (upper wall right end) at (2.8,2-1.2);
\coordinate (upper wall size) at (3.25+2.8,0.75);
\coordinate (lower wall left end) at (-3.25,-2+1.2);
\coordinate (lower wall right end) at (2.8,-2+1.2);
\coordinate (lower wall size) at (3.25+2.8,-0.75);
\draw[-, >=latex, line width=0.8] (upper wall left end) -- (upper wall right end);
\draw[-, >=latex, line width=0.8] (lower wall left end) -- (lower wall right end);
\fill[pattern=north west lines] (upper wall left end) rectangle ++(upper wall size);
\fill[pattern=north west lines] (lower wall left end) rectangle ++(lower wall size);
\draw[line width=0.7] (particle 1) circle (15pt);
\draw[line width=0.7] (particle 2) circle (15pt);
\draw[line width=1, blue] (particle 3) circle (15pt);
\draw[line width=1, blue] (particle 4) circle (15pt);
\draw[line width=0.7] (particle 5) circle (15pt);
\draw[blue] (0.4,-0.3) node[below] {\small defect!};
\draw[gray] (particle 1) node[] {\small up};
\draw[gray] (particle 2) node[] {\small down};
\draw[gray] (particle 3) node[] {\small up};
\draw[gray] (particle 4) node[] {\small up};
\draw[gray] (particle 5) node[] {\small down};
\end{tikzpicture}
\end{center}
\caption{\textbf{Loosely packed hard-disk particles in a quasi one-dimensional slit.}
For the same slit as in Fig.~\ref{fig geometry quasi 1 dim 1p5} (wall separation of $L\!=\!1.5$), 
we show a less densely packed system to illustrate a defect (in blue) with respect to the optimal zigzag structure. The probability of finding such defects is quantified by the contact 
peak of the pair distribution function $g(z_1\!=\!0.25,z_2\!=\!0.25,x\!=\!1)$.
}
\label{fig geometry quasi 1 dim v2}
\end{figure}

In Fig.~\ref{fig results quasi 1 dim g}
we show, as dotted lines, the inhomogeneous two-body correlation function $g$, as a function of  $x$ for $z_1\!=\!z_2$,  corresponding to the one-body density profiles of Fig.~\ref{fig results quasi 1 dim rho} (same shades of green). 
For reference, the exact results for $g$ obtained by Montero and Santos \cite{Q1D_santos_structure} are displayed as solid brown lines. 
The red framed panel A shows the case for $z_1\!=\!z_2\!=\!0$, which is path 1 in Fig.~\ref{fig geometry quasi 1 dim 1p5}.
As $\langle N \rangle$ is increased the packing oscillations become much more pronounced and the contact value, at $x\!=\!1$, grows rapidly.
Moreover, the second peak first shifts to the left and then piles-up from the right against $x\!=\!2$, which corresponds to three particles forming a line along the center of the channel. At the same time the value of $g$ tends to zero to the left of $x\!=\!2$ to allow the third particle to take place at a fixed position as the packing increases.
Beyond $\langle N \rangle \!=\! 0.9$ the most favorable state switches towards the zigzag state as the system `buckles' and co-linear arrangements of particles become unstable, thus much less probable. This onset of the zigzag state is associated with long-range ordering along the channel, with the oscillatory behaviour of $g$ becoming non-vanishing even for large values of $x$.
In panel A of Fig.~\ref{fig results quasi 1 dim g} the curve obtained using the inhomogeneous PY method at packing $0.9$ already exibits long-range oscillations, which tests the limits of our 
numerical algorithms.
For bigger $\langle N \rangle$-values our numerical scheme tends to break down as the oscillations 
reach the end of our numerical grid.
Nevertheless, within the range of packing for which we could obtain accurate converged PY solutions, 
namely $0\!<\!\langle N \rangle\!\le\!0.9$, the agreement with the exact reference data is excellent.
Note that for packing $0.9$ the slight deviation visible between the PY predictions and the exact data present the first, albeit still minor, indication that the inhomogeneous PY theory is, in fact, an approximation.

\begin{figure}[t!]
\includegraphics[width=\linewidth]{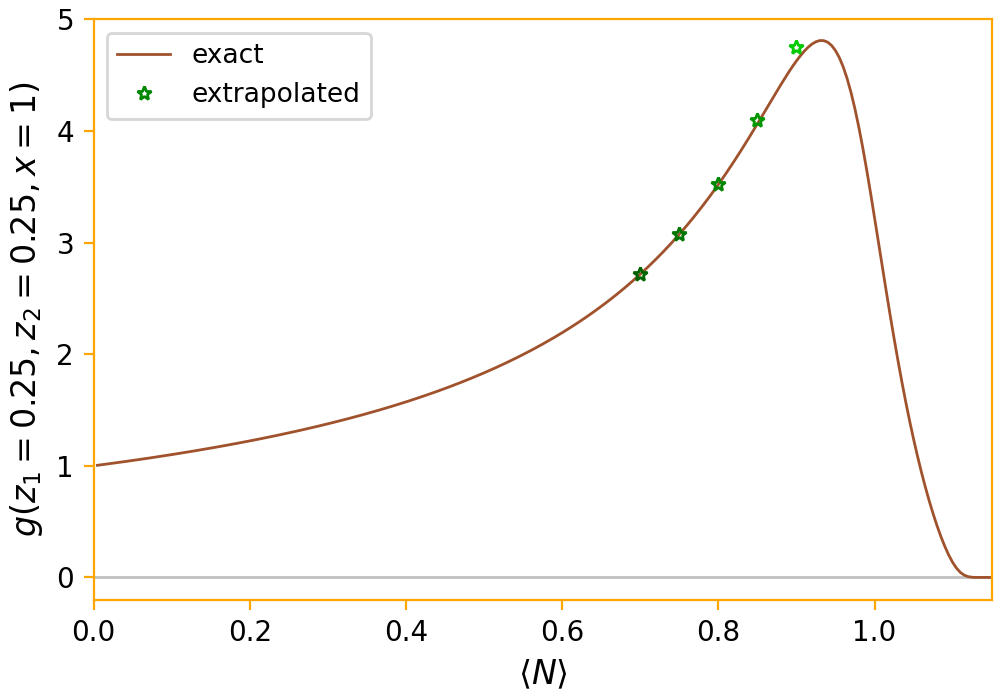}
\caption{\textbf{Defect order parameter in the quasi-1D regime.}
We show the defect order parameter, given by the value of $g$ at contact, as a function of $\langle N\rangle$. The brown solid curve is the exact result \cite{Q1D_santos_structure}.
The stars of different shades of green correspond to those already displayed in panel B of Fig.~\ref{fig results quasi 1 dim g}. The first four darker stars, for $\langle N\rangle\!=\!0.7,0.75,0.8$ and $0.85$, are very close to the exact result. The last light green star, for value $\langle N\rangle\!=\!0.9$, shows some slight deviation. 
}
\label{fig results quasi 1 dim contact peak}
\end{figure}

Panel B of Fig.~\ref{fig results quasi 1 dim g}, framed in yellow, corresponds to path 2 in Fig.~\ref{fig geometry quasi 1 dim 1p5}, with $z_1\!=\!z_2\!=\!0.25$. 
The considered density correlations are thus those between two points located at wall contact and separated by a longitudinal distance $x$. Both the accuracy and the general behaviour of $g$ along this path is similar to that shown in panel A. However, on close inspection, the periodicity of the oscillations is different. Indeed, as the packing is increased the second peak piles-up from the right against the position $x\!=\!\sqrt{3}\!=\!1.732$, shown as a dashed grey vertical line on the figure. This corresponds exactly to the distance between the centers of next-nearest neighbour particles along path 2 in the perfect packed zigzag state (see Fig.~\ref{fig geometry quasi 1 dim 1p5}). The next-nearest neighbour particles cannot approach closer than this separation length, even before perfect packing is reached.
This is distinct from the behaviour at the center of the channel (path 1), where the probability to have three particles in a line first increases as a transient state before vanishing again to allow the creation of the zigzag state, as commented previously for panel A.
In panel B of Fig.~\ref{fig results quasi 1 dim g} extrapolated contact values are indicated as green stars; indeed for technical reasons there is no grid-point at exactly $x\!=\!1$ in our inhomogeneous PY numerical scheme. Along path 2 the contact value of $g$ is of particular interest, since  
its value plays the role of a `defect order parameter' in the sense that it quantifies the probability to find two particles directly side-by-side on the `yellow line' (see Fig.~\ref{fig geometry quasi 1 dim 1p5}). This geometrical outcome is a common occurance at lower packings, but becomes impossible to achieve as the perfectly packed zigzag state is approached. This is illustrated in Fig.~\ref{fig geometry quasi 1 dim v2}.

The defect order parameter is plotted as a function of the packing in Fig.~\ref{fig results quasi 1 dim contact peak}. The stars are the same extrapolated contact values as shown in panel B of Fig.~\ref{fig results quasi 1 dim g} and the brown solid curve is the exact result due to Montero and Santos \cite{Q1D_santos_structure}.
For $\langle N\rangle\!=\!0.7,0.75,0.8$ and $0.85$, the defect order parameter agrees almost perfectly with the exact data, while, for $\langle N\rangle\!=\!0.9$, we observe some (unsurprising) slight deviation.
We note that the exact brown solid curve shows a rapid drop in value around $\langle N\rangle\!=\!1$. This indicates a change of regime, from a mostly disordered fluid to a long-range ordered zigzag state.
For packing greater than $2/\sqrt{3}$ the defect order parameter is zero, which means that the first peak of $g$ has fully vanished along path 2.
Finally, we would like to point out that the onset of the zigzag state does not constitute a true phase transition, but nevertheless occurs over a relatively small range of $\langle N\rangle$-values.

\subsubsection{The Percus-Yevick closure in quasi-1D}

The results presented above demonstrate that 
the PY closure is highly accurate for hard disks 
under quasi-1D confinement,
given the level of agreement with the analytical solution of Montero and Santos \cite{Q1D_santos_structure}. 
For the range of packings for which we can 
obtain fully converged numerical solutions, one could be forgiven for suspecting that 
the inhomogeneous PY theory may even be exact.
Only for the most demanding case we could access, 
namely $\langle N\rangle\!=\!0.9$, can slight deviations between the PY theory and the exact result be observed. 
We anticipate that these deviations would further increase at higher packings, but have confidence that the PY approximation would continue to make reliable and accurate predictions.  
However, for numerical reasons, we cannot 
yet confirm this expectation. 
The dual requirements of a fine spatial resolution, needed to capture strongly localised peaks in $g$, and a very large computational domain along the channel, required to deal with the rapid growth of long-range longitudinal order, exceed our current computational capabilities for $\langle N\rangle\!>\!0.9$. 
It is thus of interest to look at the PY closure from an analytical perspective to better understand why it works so well and what could potentially be improved.

When analyzing integral equation approximations 
it is often useful to consider their low-density Mayer cluster expansion. 
For a definitive account of this method we recommend the book chapter by Stell \cite{StellCluster}.
The leading order contributions to the \textit{exact} two-body direct correlation function 
are as follows
\begin{align}\label{exact_expansion}
     c_{\text{exact}}(\rv_1,\rv_2)
      &= \MayerA \;+\; \MayerB \;+\; \MayerC \;+\; \MayerD \\
&\,+\; \MayerE \;+\; \MayerF \;+\; \MayerG \;+\; \MayerH 
\;+\; \cdots,\notag
\end{align}
where the black circles are integration field points, weighted with a factor of the inhomogeneous one-body density, 
and the open circles are labelled root points 
corresponding to the coordinates $\rv_1$ and 
$\rv_2$. 
The lines connecting the circles represent Mayer-function bonds, where the Mayer function between coordinates labelled 
$i$ and $j$ is defined according to
\begin{equation}\label{mayer}
f(r_{ij})=e(r_{ij})-1, 
\end{equation}
with the familiar Boltzmann factor given by
\begin{equation}\label{boltzmann}
e(r_{ij}) = \exp\left(-\beta \phi(r_{ij})\right). 
\end{equation}
In contrast to the exact expansion \eqref{exact_expansion} 
the PY approximation generates \cite{mcquarrie}
\begin{align}\label{PY_expansion}
     c_{\text{PY}}(\rv_1,\rv_2)
     & = \MayerA \;+\; \MayerB \;+\; \MayerC \\
&\,+\; \MayerE \;+\; \MayerG \;+\; \MayerH 
+ \cdots.\notag
\end{align}
Thus, to quadratic order in the density,  
the difference between the exact and PY expressions is given by the sum of two diagrams
\begin{equation}\label{difference}
     \Delta c(\rv_1,\rv_2) = \MayerD + \MayerF  \;\;,
\end{equation}
where $\Delta c 
\!\equiv\!c_{\text{exact}} 
\!\!-c_{\text{PY}}$.
These two diagrams, neglected within 
the PY approximation, are special, in the sense that neither of them have a Mayer bond directly connecting the root points.
Unlike the other terms in \eqref{exact_expansion}, for purely hard-particle interactions, these 
contribute a finite `tail' to the exact direct correlation function for separations 
$r_{12}\!>\!d$, which is omitted in the PY theory 
(see equation \eqref{HD PY closure}).  
In view of the definitions \eqref{mayer} and \eqref{boltzmann} the two diagrams contributing to $\Delta c$ can be combined into a single diagram
\begin{equation}\label{tail}
     \Delta c(\rv_1,\rv_2)
     = \MayerSpecial \;\;,
\end{equation}
where the thick dotted line represents a Boltzmann-factor bond. 
While a detailed discussion of the properties of 
the diagram \eqref{tail} would go beyond the scope of the present work, we can make some general observations regarding its evaluation in confined geometry. 

For a strictly 1D system of hard rods the root points are located at positions $x_1$ and $x_2$. 
For separations $x_{12}\!>\!d$, the diagram \eqref{tail} vanishes, since the presence of the Boltzmann-factor bond between field points renders the integrand identically zero. 
Similar considerations can be applied to higher-order diagrams to prove exactly that the tail of $c$ vanishes in 1D and, hence, that the PY theory becomes exact in this limit. 
For hard disks in 2D the double integral required to evaluate \eqref{tail} runs over the overlap 
`lens' between two excluded volume disks (i.e.~disks of radius $d$) centered at $\rv_1$ and $\rv_2$, respectively. 
However, the Boltzmann-factor bond introduces the constraint on the integrand that the two field points being integrated within the lens only generate a finite contribution when they are separated by more that one particle diameter. 
In a bulk 2D system this integration generates 
a relatively small (when compared to the values of $c$ for $r_{12}\!<\!d$) amplitude tail, which is consistent with the known fact that the PY approximation is very good for bulk hard disks 
\cite{lado_disks}. 
For disks confined by hard walls to a channel 
of width $L$, the inhomogeneous one-body density, 
which is only finite within a strip of width $L\!-\!d$, must be included in the integrand as a weight factor at each field point. 
Upon entering the quasi-1D regime this strongly reduces (but not fully cancels) the number of field-point configurations 
which make a nonzero contribution to the integral in \eqref{tail}. 
(Similar considerations can be applied to the higher-order diagrams in the expansion of 
$\Delta c$.) 
The amplitude of the tail of $c$ thus 
decreases to zero as the slit width is reduced below $L\!=\!2d$ and the exact result in the 1D limit is recovered. 

The above considerations not only help to rationalize the excellent performance of the PY approximation for quasi-1D hard disks, but may also suggest a route to developing improved closures. 
If we focus on hard-repulsive particles,
there exist a number of closure schemes which are known to perform better than PY 
(e.g.~Verlet or Martynov-Sarkisov \cite{tschopp8}). All of these improved schemes are designed to treat systems in either 2D or 3D 
(usually the latter) by making some approximation to the tail of $c$. 
However, since all currently known integral equation closures 
were created with bulk systems in mind, this tail does not vanish upon dimensional reduction to 1D and so the exact (PY) result is not recovered. 
This raises the possibility of exploiting dimensional crossover 
to obtain new and more flexible closures of the 
inhomogeneous OZ equation.  
These new approximations would be constructed to preserve the desired dimensional crossover properties, while still remaining 
accurate for bulk systems.

\section{Discussion \& conclusions}

In this work we have used the inhomogeneous PY integral equation theory to study the equilibrium behaviour of hard disks confined between parallel hard walls. An interesting feature 
of this set-up is that, for quasi-1D confinement, 
the system exhibits a strongly ordered zigzag state 
at high packing. 
The onset of this state, which occurs within a relatively small range of $\langle N\rangle$, is associated with the growth of long-range correlations along the channel and the formation of a strongly inhomogeneous density profile. 
These features are not only generic for many substrate ordering phenomena (e.g.~wetting of a 
wall by the crystal phase \cite{Dijkstra}), but are also of great relevance for understanding the interface between coexisting liquid and vapour phases as well as the related effects of wetting and drying \cite{evans92,wetdry}.

Correctly capturing the feedback between the one-body density profile and the transverse two-body correlations within the interface is an important requirement for any theoretical approach 
\cite{evans92,tschopp1}.
This is explicitly realized within the present inhomogeneous integral equation theory by coupling the inhomogeneous OZ equation \eqref{Inhomogeneous OZ equation} with the LMBW sum-rule \eqref{LMBW}. 
Since the two-body correlation functions 
are functionals of the one-body density the inhomogeneous integral equation method may legitimately be regarded as an alternative form of DFT. 
Although its numerical implementation is certainly 
more demanding that the standard scheme, which is focused on the free energy functional, the integral equation approach 
has the advantage that it allows controlled approximations to be applied directly at the level of the two-body correlations. 
This enables, for example, the no-overlap core condition to be rigorously imposed in systems of hard particles, something which is very difficult 
to achieve when constructing approximations to a free energy functional, e.g.~FMT. 

A very convenient feature of the inhomogeneous integral equation method is that it can be applied not only for any external potential, but also any pair interaction potential.  
Once a closure has been chosen (e.g.~the general form for the PY approximation \eqref{PY closure}), then various pair potentials can be investigated without any modification to the theory and only minimal changes to numerical code. 
On a more fundamental level, it has recently been pointed out that the closure relation is an intrinsic property of the system 
\cite{tschopp9}, since it depends only upon the pair interaction and 
is independent of the external field. 
The intrinsic character of the closure in some sense mirrors that of the excess Helmholtz free energy functional in standard DFT. 
Approximating either of these mathematical objects 
(closure or functional) captures collective behaviour arising purely from the 
particle interactions within the model system, independently of any external influence.

Regarding future work, 
one of the clear practical challenges when using inhomogeneous integral equation methods is to obtain accurate 
numerical solutions without exhausting available numerical resources. 
While it would be highly desirable to obtain predictions from the inhomogeneous PY theory within 
the zigzag state (i.e.~for $\langle N\rangle\!>\!1$), 
our current numerical methods break down as the longitudinal correlations start to become very 
long-range. 
Addressing such issues will be important for future integral equation studies of interfacial phenomena.  
We are currently investigating the application of 
state-of-the-art pseudospectral techniques to improve both the efficiency and accuracy of our algorithms.
Finally, we note that the exact solution of Montero and Santos \cite{Q1D_santos_structure} may turn out to be valuable when seeking to develop new inhomogeneous closures beyond the PY approximation, 
since this exact solution presents a rare case for which the tail of $c$ can be determined exactly.

\bibliography{references_invited2}

\end{document}